\documentclass[entropy,article,submit,moreauthors,pdftex]{mdpi} 

\usepackage{amsmath}
\usepackage{latexsym}
\usepackage{amssymb}

\preto{\abstractkeywords}{\nolinenumbers}
\firstpage{1} 
\pubvolume{1}
\issuenum{1}
\articlenumber{0}
\pubyear{2020}
\copyrightyear{2020}
\history{Received: date; Accepted: date; Published: date}

\Title{CLUSTER STRUCTURE OF OPTIMAL SOLUTIONS IN BIPARTITIONING OF SMALL WORLDS}

\Author{Adam Lipowski $^{1,\dagger,*}$\orcidA{}, Ant\'onio L. Ferreira $^{2,\dagger}$ and Dorota Lipowska $^{3,\dagger}$}

\AuthorNames{Adam Lipowski, Antonio L. Ferreira, Dorota Lipowska}

\address{%
$^{1}$ \quad Faculty of Physics, Adam Mickiewicz University in Pozna\'n, Poland; lipowski@amu.edu.pl\\
$^{2}$ \quad Departamento de F\'{i}sica, I3N, Universidade de Aveiro,  Portugal; alf@ua.pt\\
$^{3}$ \quad Faculty of Modern Languages and Literature, Adam Mickiewicz University in Pozna\'n, Poland; lipowska@amu.edu.pl}
\corres{Correspondence: lipowski@amu.edu.pl;}

\firstnote{These authors contributed equally to this work.}

\abstract{Using a simulated annealing, we examine a bipartitioning of small worlds obtained by adding a fraction of randomly chosen links to a one-dimensional chain or a square lattice. Models defined on small worlds typically exhibit a mean-field behaviour, regardless of the underlying lattice. Our work demonstrates that the bipartitioning of small worlds  does depend on the underlying lattice. Simulations show that for one-dimensional small worlds, optimal partitions are finite size clusters for any fraction of additional links. In the two-dimensional case, we observe two regimes: when the fraction of additional links is  sufficiently small, the optimal partitions  have a stripe-like shape, which is lost for larger number of additional links as optimal partitions become disordered. Some arguments, which interpret additional links as thermal excitations and refer to the thermodynamics of Ising models, suggest a qualitatitve explanation of such a behaviour.
The histogram of overlaps  suggests that a replica symmetry is broken in a one-dimensional small world.  In the two-dimensional case,  the replica symmetry seems to hold but with some additional degeneracy of  stripe-like  partitions. 
}

\keyword{small world; graph partitioning; simulated annealing; optimization; replica symmetry}  
\begin{document}

\section{Introduction}
Optimization problems draw considerable interest of computer scientists, engineers, economists or mathematicians. Some of the optimization problems might be related with certain physical many-body problems and in such a case methodology of statistical mechanics might be used~\cite{hartmann2006}. 
Indeed, when suitable translated optimization problems manifest quenched disorder, energy barriers, or various phase transitions. Such characteristics imply interesting analogies to some glassy or magnetic  systems  and the usage of methods developed in the physical sciences such as simulated annealing or the replica technique, turns out to be remarkably successful~\cite{krzakala2016}.

A  graph bipartitioning is an optimization problem where one has to divide vertices of a graph into two classes so that to minimize the number of links between vertices of different classes. Such a problem appears in various contexts such as  VLSI circuit design~\cite{karypis}, parallel computing~\cite{pothen}, or computer vision~\cite{kolmogorov}.    Statistical mechanics approaches exploit the analogy with the Ising model and are  particularly fruitful in the random graph version of this problem. Such  a version was studied numerically using a simulated annealing~\cite{banavar,martin} or an extremal optimization~\cite{boetcher} but important analytical results were also obtained using  the replica method~\cite{fu,liao,mezard}, the technique, which was primarily developed for studying disordered systems. In a more recent  work, in which the structure of nearly optimal partitions was analyzed, some predictions concerning the replica symmetry breaking in this problem were made~\cite{percus}, which were subsequently verified using the belief propagation method~\cite{zdeborova}. The bipartition  problem was also examined for directed random graphs and it was shown that a similar replica symmetry breaking takes place~\cite{liplipferr}. 

In random graphs, links between  vertices are randomly distributed, which is often in contrast to many real networks, where vertices may be embedded in space and a probability that a link exists between a pair of vertices decreases with a distance between  these vertices. As an extreme case, one may mention regular Cartesian lattices, where links exist only between the nearest neighbours. Somewhere in between random graphs and regular lattices, one can situate the so-called small-world networks~\cite{watts}, which are drawing a considerable attention recently~\cite{smallworlds}. Small worlds  might be constructed by adding a certain amount of randomly chosen links to a regular $d$-dimensional lattice. It turns out that even a small fraction of such random, and typically long-range, links considerably affects the behaviour of models constructed on such networks.  Similarly to models on random graphs, they 	often exhibit the so-called mean-field behaviour and the underlying $d$-dimensional regular lattice often plays only a minor role~\cite{lopes}. 
Such a behaviour is especially typical to Ising-type models, but for example competition of cooperation and dishonesty \cite{capraro} or epidemic spreading  \cite{liu2015epidemics} might lead to much more reach and different behaviour. Let us notice that bipartitioning of regular Cartesian lattices is nearly trivial and results in optimal partitions being simple and compact clusters such as sections ($d=1$) or stripes ($d=2$). These simple partitions are actually ground-state configurations of the Ising model subject to the constraint of zero total magnetization.

Random graphs and regular Cartesian lattice constitute very important classes of graphs and statistical mechanics of their partitioning is already understood. At the same time these graphs are the limiting cases of the small worlds. It is perhaps interesting to ask whether partitioning of small graphs can be to understood or related with the known behaviour of these limiting cases. In our manuscript we will examine how optimal partitions change when we add  some randomly chosen bonds to the underlying regular lattice.  We  suggest that random links may act as thermal excitations, which perturb the regular-lattice partitions. Our results show that bipartitioning of small worlds can be to some extent understood by referring to a thermodynamic behaviour of the Ising model on the underlying regular network.

In section II, we describe our model and numerical method. In section III, we present the results obtained for the one- and two-dimensional small worlds. We conclude in section IV.
 
\section{Model and simulated annealing}

In the graph bipartitioning, one has to divide the graph of $N$~vertices into two classes of  equal size, here marked as~$\oplus$ and~$\ominus$, so that the partition cost, namely the number of links between vertices of opposite signs, is minimal. 

A bipartition of a graph is fairly analogous to the Ising model, in which there is a spin variable~$S_i=\pm 1$ on each vertex~$i$, and the system is described by the following Hamiltonian
\begin{equation}
H=-\sum_{(i,j)} S_iS_j.
\label{hamiltonian}
\end{equation}
In the above equation, the summation is over pairs of vertices connected by links of a graph, and the system is subject to the constraint that the numbers of~$\oplus$ and~$\ominus$ are equal, namely $\sum_{i=1}^N S_i=0$.
In terms of spin variables, the partition cost~$B$ can be written as
\begin{equation}
B=\frac{1}{2}\sum_{(i,j)} (1-S_iS_j).
\label{cost1}
\end{equation}
Finding an optimal partition becomes thus equivalent to finding the lowest energy of the ferromagnetic Ising model subject to the constraint of zero magnetization.
A number of approaches to graph bipartitioning, which exploit the above analogy to the Ising model,  were developed.

In the present paper, we analyse a bipartitioning of small worlds. To generate such graphs, we add to a regular (Cartesian) lattice $M$~links, which join two randomly chosen vertices (excluding multilple links). To find an optimal partitioning, we use a simulated annealing~\cite{kirkpatrick}. For a given graph, starting from randomly assigned spin variables {$S_i$}, the algorithm  selects a pair with opposite values and exchanges them according to the Metropolis update, namely with probability min($1,\exp(-\Delta B/T)$), where $\Delta B$ is the change of the cost. During the run, the temperature-like parameter $T$ is reduced as $T=T_0\exp(-rt)$, where $r$ is the cooling rate and and $t$ is the simulation time (a unit of time is defined as an update of $N$~pairs of vertices). We used $T_0=1$ and $r=10^{-5}$, but to increase the accuracy of our protocol, we made several such annealings for a given graph  ($\sim 100$, each starting from a different initial spin configuration) and selected the final configuration with the lowest value of the partition cost~$B$. We examined the structure of such (nearly) optimal solutions and calculated the average partition cost, where averaging was over independently generated graphs with the given values of~$N$ and~$M$. 

Furthermore, we examined the so-called replica symmetry. This symmetry is related to the similarity of different ground-state configurations. In a replica-symmetric phase, such configurations are to a large extent similar, while in a replica-symmetry-broken phase, they are much different. This symmetry has been extensively studied in the hope of clarifying the nature of the ordering in spin glasses~\cite{marinari,young} as well as in various optimization contexts~\cite{krzakala,monasson}. To examine the symmetry, we generate a graph and run our simulated annealing protocol, which finds two replicas~$\mathcal{A}$ and~$\mathcal{B}$. Assuming that at the end of the run these replicas are specified by their spin configurations $\{S_i^\mathcal{A}\}$ and $\{S_i^\mathcal{B}\}$, respectively, we calculate the overlap~$q$ defined as follows
\begin{equation}
q=1/N\sum_{i=1}^N S_i^\mathcal{A}S_i^\mathcal{B}.
\label{overlap}
\end{equation}
To calculate~$q$ for a given graph, we use $10^2$ pairs of replicas and  then we also average over $10^3$ different graphs.

Although simulated annealing is a general purpose optimization technique that was successfully used in numerous applications, its accuracy is hard to estimate. Nevertheless, we hope that comparing  numerical results  for graphs of different size $N$ and different cooling protocols ($T_0,\ r$) we were able to draw some plausible conclusions.

\section{Results}
 In the following, we present the results of our calculations obtained for small worlds on a linear chain ($d=1$) and a square lattice ($d=2$).
We expect that the simulated annealing that we used gets less accurate with increasing the size of the graph~$N$. This may be particularly important in  calculations of the overlap~$q$. To have a similar accuracy, we made  calculations   for graphs of the same size~$N$ for both $d=1$ and $d=2$.
\subsection{d=1}

\begin{figure}
\centering
\includegraphics[width=0.5\columnwidth]{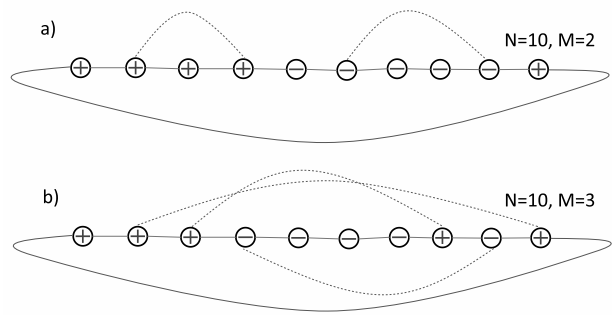}
\vspace{0mm}
\caption{a) When the number of additional links (dashed lines) is small,  the cost of the optimal configuration composed of two clusters of length $N/2$ is  $B=2$.  Note the periodic boundary conditions. b) For larger number of additional links, an optimal configuration composed of smaller clusters has the cost $B=4$.}
\label{graphd1}
\end{figure}

First, let us consider a linear chain of size~$N$ without additional links ($M=0$). In this case, the optimal partition, the cost of which is $B=2$, consists of two clusters  of length~$N/2$ (within each cluster spin variables take the same values). Such a two-cluster partition may be optimal  even when $M$ is positive and ''not too big'' (Fig.~\ref{graphd1}a).  

\begin{figure}
\centering
\includegraphics[width=0.5\columnwidth]{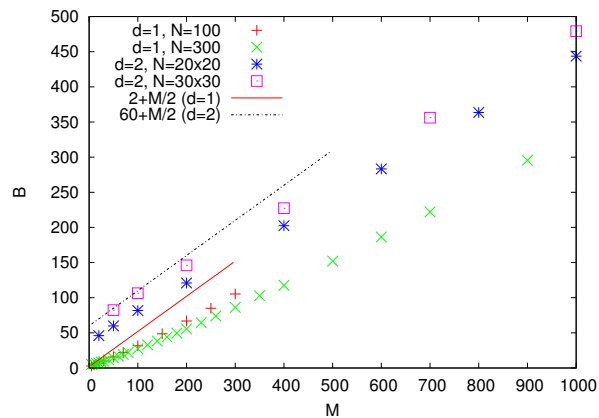}
\vspace{0mm}
\caption{The average partition cost~$B$ determined using simulated annealing as a function of the number of additional links~$M$. Straight lines correspond to the two-cluster estimation, where half of the additional links contribute to the partition cost.}
\label{cost}
\end{figure}

For larger~$M$, optimal partitions typically consist of several smaller clusters (Fig.~\ref{graphd1}b) and their average cost increases with~$M$.   Let us notice that a two-cluster partition may serve as a simple approximate estimation of the partition cost~$B$. Indeed, assuming a random position of such a partition, we may expect that on average half of additional links (which connect randomly chosen nodes)  connect vertices of opposite signs. Thus,  the average cost of such a partition is \mbox{$B=2+M/2$}. Not surprisingly, the partition cost as determined  using simulated annealing is smaller than this estimation (Fig.~\ref{cost}).

\begin{figure}\centering\includegraphics[width=0.5\columnwidth]{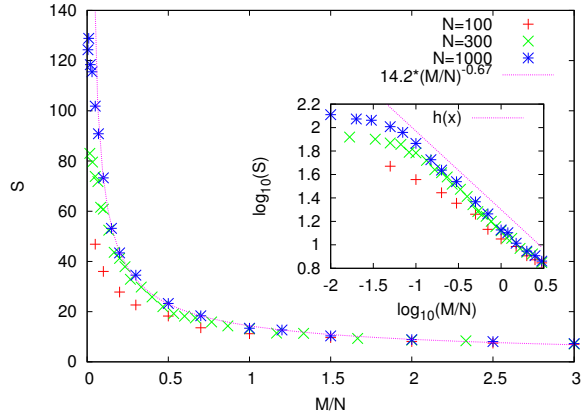}
\vspace{0mm}
\caption{The average cluster size~$S$ as a function of~$M/N$. A power-law diverging fit to our data (line) for $N=1000$ and $M/N>0.1$ suggests that the possible divergence of~$S$ takes place at $M/N=0$. Hence, for any $M/N>0$, the optimal solution consists of finite size clusters. Inset shows our data on the log-log scale and the dotted straight line has a slope -0.67. For small $M/N$ we observe a deviation from the power-law behaviour and we attribute it to finite-size effects.}
\label{avclustersize}
\end{figure}

To examine in more detail the structure of optimal partitions, we calculated the average size~$S$ of (for example) $\oplus$-clusters. We adapt a usual percolation theory definition~\cite{percol} of the average cluster size. If the $\oplus$~spins in the optimal partition form clusters of size $s_1,\ s_2,\ldots,\ s_k$,  we calculate the average cluster size as $S=2/N\sum_{i=1}^k s_i^2$. For example, for the partition in Fig.~\ref{graphd1}b, we obtain $S=\frac{2}{10}(4^2+1^2)=17/5$ (note periodic boundary conditions). To calculate the average cluster size~$S$ for the given values of~$N$ and~$M$, we averaged it over $10^2$ independently generated graphs. Our numerical results  show that $S$ is a decreasing function of~$M$ (Fig.~\ref{avclustersize}). Morover, the data for different~$N$ plotted as a function of~$M/N$ seem to collapse on a single curve, which indicates that the relevant parameter is actually the density of additional links. Although one can notice strong finite-size effects (for small $M/N$), the numerical  data  suggest that $S$~diverges upon approaching $M/N=0$. Our data for $N=1000$ and $M/N>0.1$ are well fitted with  a power-law function that diverges at $M/N=0$. It suggests that for any $M/N>0$, the optimal partition consists of finite size clusters.

It is interesting to ask how many optimal partitions exist for a given graph. Of course, a global up-down symmetry ($S_i\rightarrow -S_i$) of  Hamiltonian (\ref{hamiltonian}) implies that there are at least two such partitions. However, it can be speculated that,  in principle, there may be more of optimal partitions not related by any symmetry. The double degenerate scenario is usually referred to as replica symmetric and when there are more of optimal partitions, the replica symmetry is broken. Actually, closely related problems appear in various glassy or disordered systems and sophisticated techniques were used to address them~\cite{krzakala2016}. To examine this problem, we calculated the overlap~$q$ as defined in Eq.~(\ref{overlap}).
On general grounds, one expects that in the replica symmetric regime, the distribution of~$q$ is strongly peaked at a value close to $q=\pm 1$, which corresponds to a double-degenerate-valley structure of the ground state. In the replica broken-symmetry phase, much broader distribution is expected, which even at $q=0$ may remain positive.

\begin{figure}\centering\includegraphics[width=0.5\columnwidth]{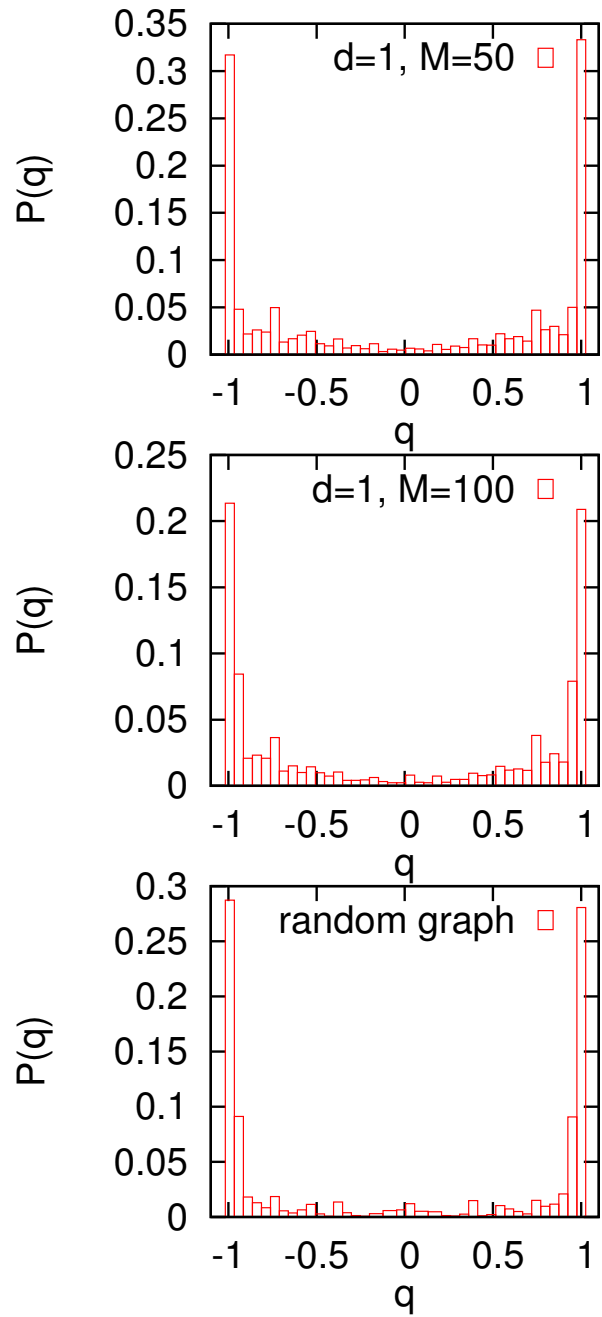}
\vspace{0mm}
\caption{The probability distribution~$P(q)$ of the overlap~$q$. The calculations were made for the $d=1$ small world with $N=100, \ M=50$ (upper panel), $N=100, M=100$ (middle) and random graph with $N=100$, $z=8$ (bottom). A small but nonzero value of~$P(q)$ at $q=0$ might indicate that in all cases the replica symmetry is broken. In the case of random graphs, there are some independent arguments and calculations that support such a claim~\cite{percus,zdeborova}.}
\label{overlapd1}
\end{figure}

The calculation of the histogram~$P(q)$ is usually a very demanding computational task and the results are sometimes difficult to interpret. Our calculations for $N=100$, and $M=50$ and~100 show pronounced peaks around $q=\pm 1$, but the distributions $P(q)$ are quite broad with a small  value at $q=0$ (Fig.~\ref{overlapd1}). This indicates that for a given graph, there is a certain (albeit small) probability that the two optimal partitions that are found using  the simulated annealing are totally independent. For comparison,  we also present the results of the calculations for the Erd\H{o}s--R\'enyi random graph with the average vertex degree $z=8$ obtained using the same numerical procedure (Fig.~\ref{overlapd1}, bottom panel). In this case, the bipartitioning is known to be in the replica-symmetry-broken regime~\cite{percus, zdeborova}. Our results for the random graph look similar to the small world data except for a slightly more pronounced peak at $q=0$. The numerical data do not provide a strong evidence, but in our opinion they suggest that in the $d=1$ small-world model, the replica symmetry is broken, at least for the examined values of $M/N$.  

\subsection{d=2}
We also analysed two-dimensional small worlds obtained by adding $M$ randomly chosen links to a square lattice of the linear size~$L$ ($N=L^2$). Similarly to the $d=1$ case, when $M$ is small, the optimal partition consists of two stripes of the width~$L/2$. For $M=0$, the cost of such a partition is $B=2L$. For increasing $M$, the partition cost also increases (Fig.~\ref{cost}). Similarly to the one-dimensional small worlds, when $M$ is small, we may expect that a randomly placed two-stripe partition  provides a certain approximate solution, and the average cost of such a partition equals $2L+M/2$. One can notice that for $L=30$ and $M \leq 100$, the agreement with simulated annealing results is quite good (Fig.~\ref{cost}).

\begin{figure}
\centering
\vspace{-0mm}
\includegraphics[width=0.5\columnwidth]{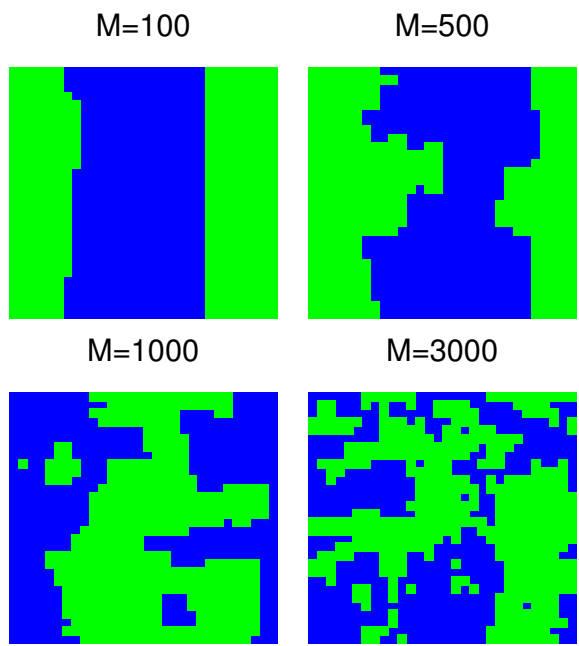}
\vspace{0mm}
\caption{Exemplary optimal configurations for a two-dimensional 30$\times$30 graph. Upon increasing the number of additional links, around $M=1000$, the stripe-like solutions turn into disordered clusters.}
\label{config}
\end{figure}

Of course, for increasing~$M$, the shape of optimal partitions changes. Namely, it may be profitable to increase the length of the boundary of stripes (which increases the cost~$B$) but to satisfy in such a way some of the additional links (which decreases the cost~$B$). The shape of some typical configurations for the $30\times 30$ system as found   using simulated annealing is shown in Fig.~\ref{config}. One can notice that a stripe-like pattern persists approximately up to $M=1000$, and for greater $M$,  optimal partitions are disordered. In view of these exemplary configurations, it is tempting to consider the additional links as generating some kind of noise into the two-stripe structure, similarly perhaps to a thermal agitation in the Ising model.

\begin{figure}\centering\includegraphics[width=0.5\columnwidth]{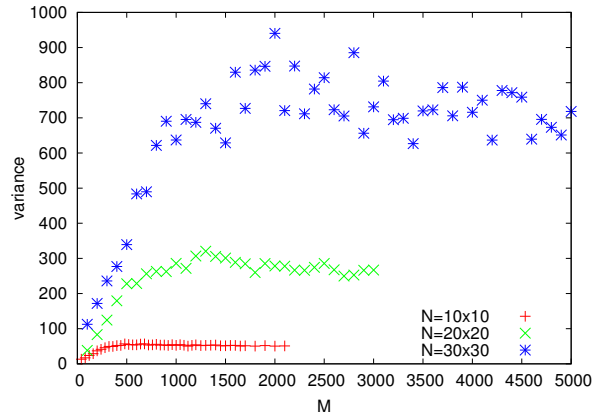}
\vspace{0mm}
\caption{The variance of the length of boundaries between clusters as a function of~$M$.}
\label{variance}
\end{figure}

To further analyse the change of shape of optimal partitions, we calculated the length of boundaries separating positive and negative clusters (which is the number of edges in the square lattice linking the~$\oplus$ and~$\ominus$ spins). We do not present numerical data but, as expected, this length is a rather smoothly increasing function of~$M$. Perhaps more interesting is the variance of this length, which after an initial  increase becomes nearly independent of~$M$ (Fig.~\ref{variance}).   
For $L=30$, the transition between these two regimes  takes place around $M=1000$, which is also the value, where  stripe-like partitions change into disordered ones (Fig.~\ref{config}).
The data in  Fig.~\ref{variance} show that for $L=10$, the transition between these two regimes  takes place around $M=300$ and for $L=20$, it is around $M=650$. Thus, approximately, the location of the transition point seems to scale linearly with~$L$ but we cannot provide an explanation of such a behaviour.

Our results presented in Fig.~\ref{config} and Fig.~\ref{variance} suggest that the model for $d=2$ has two regimes. In the first regime (for $L=30$, it corresponds to  $M\lesssim 1000$), optimal  partitions have a stripe-like stucture and the variance of the total length of boundaries increases with~$M$. In the second regime (for $L=30$, it corresponds to  $M\gtrsim 1000$), optimal partitions are disordered and the variance of the total length of boundaries is nearly constant as a function of~$M$. In our opinion, such a behaviour resembles the behaviour of the two-dimensional Ising model (e.g., on a square lattice), which remains ferromagnetic at low temperature  (first regime) and is paramagnetic at high temperature (second regime). More precisely, it would be an Ising model with a conservative dynamics and the constraint of zero total magnetization. In such a case, the ferromagnetic phase corresponds to the phase separation. In this analogy, additional links  play the role of thermal excitations and an increasing~$M$ corresponds to the increase in temperature. Let us notice that such an analogy helps us to understand the behaviour of the $d=1$ version of our model. As it is well known, the one-dimensional Ising  model (with short-range interactions only) remains paramagnetic at any positive temperature~\cite{huang}. Thus, an arbitrarily small $M/N>0$ should be sufficient to destroy the phase separation  and lead to optimal partitions being finite clusters. However, the interpretation of additional links as thermal excitations should be taken with some care. While additional links in Fig.~\ref{graphd1}b might be interpreted in such a way, those in Fig.~\ref{graphd1}a cannot be (additional links in Fig.\ref{graphd1} happen to link spins of the same orientation; in general this does not have to be the case.).

\begin{figure}\centering\includegraphics[width=0.5\columnwidth]{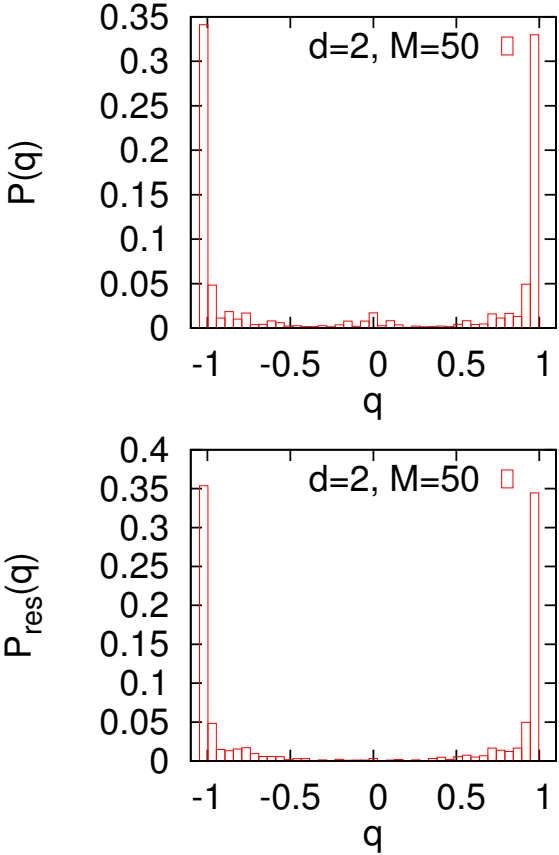}
\vspace{0mm}
\caption{The probability distributions $P(q)$ and $P_{res}(q)$.  Calculations were made for the $d=2$ small world with $L=10$ and $M=50$.}
\label{overlapd50}
\end{figure}

We also calculated the overlap distributions~$P(q)$ and the results are shown in Fig.~\ref{overlapd50}. For $L=10$ and $M=50$, one can notice a peak at $q=0$, which could indicate a replica-symmetry-broken regime . In disordered or glassy systems, replica-symmetry breaking is usually related to the formation of multi-valley structure of the configuration space. In our case, replica-symmetry breaking is related to an additional degeneracy, namely, to the fact that in the stripe-like regime, at least for certain graphs, optimal stripes can run both horizontally and vertically. Such pairs will have the overlap $q\approx 0$ and this would explain the small peak of $P(q)$ in Fig.~\ref{overlapd50}. To confirm such a scenario, we examined for each optimal partition the values of spins at the boundaries, which enabled us to clasify it as a horizontal or vertical configuration. Then, we calculated the overlaps $P_{res}(q)$, where the average is restricted to only horizontal or only vertical pairs of optimal partitions. The numerical calculations show that $P_{res}(q)$  for~$q$ close to~0 takes negligibly small values (Fig.~\ref{overlapd50}). Strong peaks at $q=\pm 1$ show that if we restrict the analysis, e.q., to horizontal configurations only, then the system remains in a replica-symmetric regime and the small peak at~$P(q)$ comes from an additional horizontal-vertical degeneracy. 

\begin{figure}\centering\includegraphics[width=0.5\columnwidth]{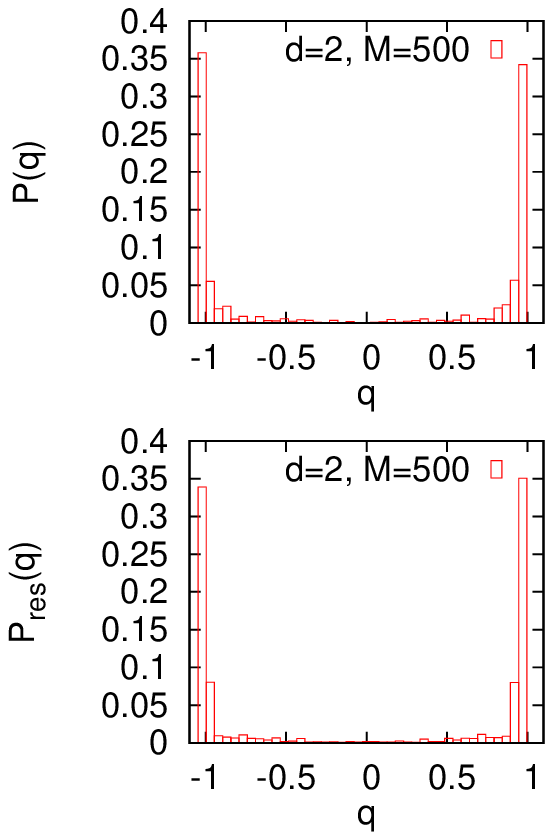}
\vspace{0mm}
\caption{Th probability distributions $P(q)$ and $P_{res}(q)$.  Calculations were made for the $d=2$ small world with $L=10$ and $M=500$.}
\label{overlapd500}
\end{figure}

We repeated the calculations for $L=10$ and $M=500$ (Fig.~\ref{overlapd500}). In this case, optimal partitions loose a stripe-like shape and, not surprisingly, $P(q)$ and $P_{res}(q)$ look almost the same. Negligibly small values at $q=0$ and strong peaks at $q=\pm 1$ suggest an (ordinary) replica-symmetric regime. It seems that for larger~$M$,  random links dominate rendering the small-world network more similar to a random graph (with a large average vertex degree), and we expect the replica symmetry to be broken~\cite{percus,zdeborova}. It may be difficult, however, to provide a convincing numerical confirmation of such a behaviour.

\section{Conclusions}
In the present paper we examined the bipartitioning of small worlds.  We analysed small worlds obtained by adding some randomly chosen links to the underlying lattice being a one-dimensional chain or a two-dimensional square lattice. For the one-dimensional chain, our results show that the optimal partitions are composed of finite size clusters for any positive fraction of additional bonds. In the two-dimensional case, when the fraction of additional links is sufficiently small, the optimal partitions have a stripe-like shape. For larger number of links, they become disordered. We suggest that random links added to the underlying regular lattice act as some kind of thermal excitation, which disturbs the compact optimal partitions. Under such interpretation, we can understand the difference between a bipartitioning of one- and two-dimensional small worlds referring simply to the thermodynamics of the Ising model on regular one- and two-dimensional lattices. Of course, the suggested association between the bipartition and the thermodynamics of the Ising model is only intuitive, if not vague, and it would be certainly desirable to provide more precise arguments. Let us also notice that the models on small worlds, due to long-range links, are generally thought to belong to the mean-field universality class~\cite{lopes}. Our work shows that the dimensionality of the underlying lattice also plays an important role in bipartitioning. 

We have also analysed the replica symmetry of optimal bipartitions of small worlds.  For the one-dimensional underlying lattice, most likely the system exhibits the replica-symmetry breaking. It is possible that such a behaviour appears for any number of additional links. Indeed, let us notice that without additional links, the replica symmetry is trivially broken (any position of a $\oplus$-cluster is allowed). Moreover, for a large number of additional links, the small world becomes similar to  a random graph with a large vertex degree, and in such a case, the replica symmetry is also known to be broken~\cite{percus,zdeborova}.  The two-dimensional case is perhaps more interesting. For a small number of additional links, our simulations show that the replica symmetry is broken but such a behaviour is related to a vertical-horizontal degeneracy of possible orientations of optimal partitions. For a larger number of additional links, optimal partitions loose a stripe-like shape, the degeneracy is removed and the model is replica symmetric. Whether this symmetry would break down for even larger number of additional links, when the small worlds would be more similar to random graphs, remains an open question yet. An additional analysis of the replica symmetry using, e.g., a message passing algorithm~\cite{zdeborova} would be certainly desirable.



\vspace{6pt}

\authorcontributions{conceptualization, A.L. and A.L.F.; methodology, A.L. and A.L.F.; software, A.L. and D.L.; validation, A.L., A.L.F., and D.L.; investigation, A.L., A.L.F., and D.L..; writing--review and editing, A.L., A.L.F., and D.L.; visualization, A.L. All authors have read and agreed to the published version of the manuscript.'', please turn to the  \href{http://img.mdpi.org/data/contributor-role-instruction.pdf}{CRediT taxonomy} for the term explanation. Authorship must be limited to those who have contributed substantially to the work reported.}

\conflictsofinterest{The authors declare no conflict of interest.} 
\reftitle{References}

\externalbibliography{yes}
\bibliography{abbrev_titles,biblio}



\end{document}